\documentclass[a4paper,12pt]{article}
\usepackage{amsmath}
\usepackage{amsfonts}
\usepackage{amssymb}
\usepackage{enumerate}
\usepackage{xcolor}
\usepackage{authblk}\numberwithin{equation}{section}

\title{Novel Edge States in Self-Dual Gravity }

\author[1]{A. P. Balachandran\thanks{balachandran38@gmail.com}}
\author[2]{ Amilcar R. de Quieroz\thanks{amilcarq@unb.br}}
\author[3]{ M.Arshad Momen\thanks{arshad@iub.edu.bd, amomen@du.ac.bd}  }
\affil[1]{Physics Department, Syracuse University, Syracuse, New York, U.S.A.}
\affil[2]{Instituto de Fisica, Universidade de
Brasilia, \\Caixa Postal 04455, 70919-970, Brasilia, DF, Brazil}

\affil[3]{Department of Physical Sciences, 
Independent University,
Bashundhara R/A, Dhaka-1212, Bangladesh\footnote{On leave of absence from Theoretical Physics Department, 
University of Dhaka, Dhaka, Bangladesh. }
}

\date{\today}

\begin{document}
\newcommand{\be}{\begin{equation}}
\newcommand{\ee}{\end{equation}}
\newcommand{\bea}{\begin{eqnarray}}
\newcommand{\eea}{\end{eqnarray}}
\newcommand{\Tr} {\mathrm{Tr~}}
\maketitle

\begin{abstract}
In contrast to the Einstein-Hilbert action, the action for self-dual gravity contains vierbeins. 
They are eleminated at the level of observables by an $SL(2,\mathbb{C})$ gauge condition implied 
by the action. We argue that despite this condition, new ``edge'' or superselected state vectors 
corresponding to maps of the spheres $S^2_{\infty}$ at infinity to $SL(2, \mathbb{C})$ arise. They 
are characterised by new quantum numbers and they lead to mixed states. For black holes, they arise
both at the horizon and the spatial infinity and  may be relevant for the black hole information paradox. 
Similar comments can be made about the Einstein-Palatini action which uses vierbeins.  

\end{abstract}
\baselineskip=18pt
\section{Introduction}

There are several systems with more than one Lagrangian for given equations of motion, but when quantised
lead to inequivalent quantum systems \cite{Book_1}. They do not differ by total derivatives of functions. We call 
these Lagrangians as {\em weakly} inequivalent Lagrangians. Simple examples of weakly inequivalent 
Lagrangians  on $\mathbb{R}^N$ are:
\be
 L_M = \frac{1}{2} \dot{x}^i M_{ij} \dot{x}^j , \quad \dot{x}^i \equiv \frac{dx^i}{dt}, \quad
 M \in \mathrm{Mat}_{N\times N} ( \mathbb{R}),\quad  \det(M) \neq 0 , \qquad M^T = M
 \label{eem}
\ee
 for different choices of the non-singular symmetric matrix  $M$. Here $x = ( x^1, x^2, \cdots, x^N) $ are the coordinates. 
  The matrix $M$  is not varied for finding the equations of motion (\ref{eem}). The latter are independent of $M$, 
  \be
  M \frac{d^2 x^i}{dt^2 } = 0 \Rightarrow  \frac{d^2 x^i}{dt^2}=0 
  \ee
 $M$ being non-singular. 
 
 For $M = m \mathbb{I}_{N \times N}, m\neq 0$, $L_M$ has the symmetry $ x \rightarrow R x$ , $ R \in SO(N)$ 
 but that is not the case for generic $M$. Also the canonical momentum $p_i = M_{ij} \dot{x}^j $ depends on $M$. 
 For such reasons, quantum theories for different $M$ mutually differ, although that is not the case with 
the equations of motion. 
 
 But there are also familiar Lagrangians in field theories which share the same symmetries and which are regarded 
 as equivalent. Examples are the Einstein-Hilbert, the Einstein-Palatini and the self-dual theory of the Einstein gravity. 
 All  of these actions are diffeomorphism invariant, but the last two actions have redundant variables in the form of vierbein variables. 
 The Lagrangian formalism gives rise to first class Gauss law constraints which help us in the elimination of these  redundant variables.

 Thus the Gauss law is central to establish some sort of equivalence between these Lagrangians. A naive application
 of Gauss law does suggest this equivalence. But the Gauss law operator is also a distribution, and therefore must be smeared 
 with test functions. When this aspect is also considered, one finds  leftover quantum state vectors associated with spatial
 infinity carrying new quantum numbers. We have discussed them in various contexts before \cite{Bal_2} and we describe
 them again here, in section 2. In  the framework of   local quantum physics, they would be associated with superselection rules. 
 We apply the methods described in  this section  to  self-dual gravity and show how edge states on  the  sphere at spatial infinity, $S^2_\infty$
 arise from the  Gauss law. 
 
 Similar remarks can be made about the Einstein-Palatini action.
 
 The conclusion is that actions based on vierbeins are not equivalent to the Einstein-Hilbert action as they contain 
 additional edge states. The discussion brings out that edge states are to be expected associated with every 
 spatial boundary, such as the black hole horizon \cite{Bal_Chandar}.  
 
 Section 3 briefly discusses the known material in the theory of self-dual action. It has a Gauss law based on the 
 non-compact group $SL(2, \mathbb{C})$. Its edge states, emergent from this Gauss law, are discussed in section 4. 
 In sections 5 and 6, it is argued that if the observables of this action are to coincide with those of the Einstein-Hilbert 
 action, the above state vectors must lead to {\em mixed} states. This is analogous to the results of Balachandran, de Queiroz and Vaidya \cite{Mix} that 
 the colored states in QCD are all mixed. 
 
 These remarks are also readily adapted to the Einstein-Palatini action.
 
{ \it Important Remark:} There is recent extensive important work on edge 
 states and low energy theorems by Strominger et al\cite{Strominger}. 
 We believe that the present work has minimal overlap with theirs.
 
 \section{The Gauss Law and Edge States in QED}
 
 Some of the important  points of interest in the treatment of Gauss law are already brought out by QED. 
 
 At a fixed time, if $E^i$ is 
 the electric field and $J_0$ the charge density, the Gauss law classically is 
 \be
  \partial_i E^i (x)  + J_0 (x) =0. 
  \label{gauss}
 \ee

 In quantum theory, following Dirac, (\ref{gauss})  is regarded as a condition on allowed states vectors $|\cdot \rangle$ : 
  \be
  (\partial_i E^i   + J_0)  (x) | \cdot \rangle =0. 
  \label{gauss_2}
 \ee
  We can regard (\ref{gauss_2}) as a condition which picks out the domain of the observables. 
  
  But $E^i$ in quantum theory are distributions.  Thus, we must integrate (\ref{gauss_2}) after multiplying by suitable test functions and rewrite it so that 
  derivatives appear only on the test functions. That is because derivatives of distributions are understood dually, in terms of 
  test functions. 
  
  So let $C^\infty$ denote infinitely differentiable functions , and $C_0^\infty \subset C^\infty$, those functions which also vanish fast at 
  spatial infinity.  Then we can rewrite (\ref{gauss_2}) 
  as 
  \be
  G(\Lambda) |\cdot \rangle := \int d^3x~ \left[ -E^i \partial_i  \Lambda  + \Lambda (x) J_0(x)\right] |\cdot \rangle =0 ,
  \label{gauss_3}
  \ee
with     $\Lambda \in C_0^\infty$. Classically, (\ref{gauss_3}) implies (\ref{gauss_2}) by partial integration since $\Lambda$ vanishes fast at infinity. 
    
    The exponentiation of  $G(\Lambda)$, namely , $e^{i G(\Lambda)}$ , generates a group.  The electron field $\Psi_0$ under its action acquires a local phase $e^{i \Lambda(x)}$
    at $x$, 
    \be
    e^{iG(\Lambda)} \Psi_0(x) e^{-iG(\Lambda)} = e^{i \Lambda(x)} \Psi_0(x)
    \label{phase1}
    \ee
    and the phase becomes $\mathbb{I}$ at $\infty$,  since $\Lambda(x) \rightarrow 0$  as $|\mathbf{x} | \rightarrow \infty$. Thus the group associated
    with $G(\Lambda)$ can be thought of as the group $\mathcal{G}_0^\infty$ of  maps from $\mathbb{R}^3$ to $U(1)$ which becomes identity at $\infty$, 
    the superscript  $\infty $ denoting this fact  while the subscript $0$ denoting that it is connected to identity. 
    
    We note that all ``observables'' are required to commute with only with these ``small'' gauge transformations $e^{i G(\Lambda)}$ . The electron field
    $\Psi_0$ is not invariant under small gauge transformations. In any case, $\Psi_0$ changes sign under $2\pi$-rotations and thus
    it is not an observable. 
    
    Reverting to (\ref{gauss_3}), we can next consider 
    \be
    Q(\mu) = \int d^3x~ \left[ -E^i ~\partial_i  \mu  + \mu (x) J_0(x)\right] 
    \ee
    where $\mu$ approaches a constant $\mu_0$ at $\infty$. 
    Let us call the group generated by  such $Q(\mu)$  as $\mathcal{G}_0$.

    For $\mu \neq \Lambda$ , $Q(\mu) | \cdot \rangle $ need not be zero. If $\mu = \Lambda$, one has $Q(\mu) = G( \mu) $ and thus 
    $Q(\Lambda) | \cdot \rangle =0$. This is the subgroup $\mathcal{G}^\infty_0 = \left\{ e^{i Q(\Lambda)}\right\}$ of $\mathcal{G}_0$  which acts
    trivially on $|\cdot \rangle$.  Thus the effective group acting on $|\cdot \rangle $ is $ \mathcal{G}_0 / \mathcal{G}_0^\infty$. 
    The group $\mathcal{G}_0 / \mathcal{G}_0^\infty$ is isomorphic to $U(1)$. For, any element of this quotient is entirely characterized 
    by  the value of $\mu$ at spatial infinity. 
    
    As an example, choose  for  $\mu$  a  globally constant $\mu = \mu_0$.  In that case, 
    
    \be
    Q(\mu) = \mu_0 \int d^3x~  J_0(x)  = \mu_0 Q( \mathbb{I}) \equiv \mu_0 Q_0 
    \label{charge_q}
    \ee
  where $\mathbb{I}$ is the constant function with the value $1$ and $Q_0$ is the canonically normalised
  charge. 
  \\
  
 An important question now is:  how do we create sectors with $Q(\mu)\neq 0$ from the vacuum? 
 For this purpose, we can consider 
  \be
  W(x) = \exp{ i \int_x^\infty dx^\lambda ~ A_\lambda (x)}
  \label{Wil_line}
  \ee
where the integral is at a fixed time  along the spacelike straight  line 
\be
\{ x +  \hat{n} l , 0 \leq  l < \infty\}
    \ee
    in the direction of a unit vector, say $\hat{n}$. Under a gauge transformation, 
    \be
    A_\lambda \rightarrow A_\lambda + \partial_\lambda \mu 
    \ee
    one has 
    \be
    W(x) \rightarrow  e^{i \mu_\infty ( \hat{n}   )}  W(x)  e^{-i \mu(x)}, 
    \ee
    where 
    \be
    \mu_\infty(\hat{n}) = \lim_{l \rightarrow \infty } \mu( x + \hat{n} l ) 
    \label{angle_inf}
    \ee
    Thus, in view of (\ref{phase1}) and  as Dirac observed \cite{Dirac},\cite{Mandelstam}, 
    \be
    W(x) \Psi_0(x) = \Psi(x)
    \ee
    is invariant under small gauge transformations whereas under large ones, 
    \be 
    e^{i Q(\mu)}  \Psi(x)  e^{-iQ(\mu)} = e^{ i  \mu_\infty ( \hat{n} ) } \Psi(x)
   \label{sky}
    \ee
    Therefore, if $\mu_\infty$ is equal to a constant $\mu_0$, the state $\Psi(x)| 0 \rangle $ , where 
    $| 0 \rangle$ is the vacuum, is the vector with $Q_0=1$. But at the same time, it is constant under small gauge transformations.

    In  $\mathcal{G}_0$,we can let $\mu _\infty$ be an angle-dependent function. We can generate such elements of $\mathcal{G}_0$
    by letting $\mu(x)$ approach an angle dependent   limit $\mu_\infty( \hat{n})$, as 
     indicated by (\ref{angle_inf}). Then on exponentiation, and restricting $\hat{n}$ to a Cauchy slice, 
     we get the Sky group of \cite{Mix} isomorphic to the maps from 
     the celestial sphere $S^2_\infty$ at infinity to U(1).

     As $\mu$ runs over its possibilities, we get a collection of Sky charges $\mu_\infty (\hat{n})$ where $\hat{n}$ now  fixes the
     representation of $\mathcal{G}_0 / \mathcal{G}_0^\infty$. We have in this case, if $Q(\mu) | \cdot \rangle=0$, 
     \be
     Q(\mu) \Psi(x) |0 \rangle = \mu_\infty ( \hat{n}) \Psi(x) | 0 \rangle
     \label{rep_sky}
    \ee
    
    Since the representation given by  (\ref{rep_sky}) depends only on the asymptotic data $\mu_\infty(\hat{n})$, 
    the corresponding quantum states are {\it edge states}.

    In the discussion above, we have assumed that the charge of the electron is unity in suitable units. For a field of 
    charge $q$, $A_\lambda$ must multiplied by $q$ while we can also easily restore $q$ appropriately in the above equations.  
    
    { \it Important Remarks} :  In previous work  \cite{Bal_QCD}  (  In particular, see the discussion in the  appendix),
     the in-state was constructed using $\hat{n}$  as a time-like unit vector. We can do that here also. 
     But then the commutators of $A_\lambda$ with other fields are known only after solving field equations 
     whereas by choosing $x+ \hat{n}l$  to lie on a Cauchy slice, we know all the equal time commutators 
     including those of $W$ and $\Psi$. 
     For this reason, in what follows, we choose $\hat{n}$ to be spacelike and be a unit tangent to a spacelike surface.

    \subsection{ On Gauge Redundancies in QED} 
    The fields in the QED Lagrangians are connections $A_\mu$ and the electron field $\Psi_0(x)$.
    Gauge invariance implies that there are redundant degrees of freedom in the QED Lagrangian.

    The new quantum state vectors which may arise from these redundant degrees of freedom get restricted by the Gauss law. 
    They do not get entirely eliminated however. The charged vectors with charges $q$ and the Sky group vectors characterised by 
    $\mu( \hat{n}) $ are examples. 
   
   Thus the redundant degrees of freedom in the Lagrangian leave an imprint on the state vectors. It is important that the redundant 
   degrees  of freedom implicit in the Lagrangian do not appear at the level of local observables \cite{Bal_2, Mix}. They must by definition 
   commute with $G(\Lambda)$, but because of the locality, they then commute also with $Q(\mu)$. The reasoning is as follows. 
   
   Let $K$ be a compact spatial region and let $\mathcal{O}_K$ be a local observable localised in $K$. Then by (\ref{gauss_3}), and its generalisation 
   to local observables 
   \be
   \left[ Q(\mu), \mathcal{O}_K \right] = \left[ Q(\mu|_K), \mathcal{O}_K \right]
   \ee
  where $\mu|_K$ = $\mu$ restricted to $K$.  But we can find a $\Lambda \in C^\infty_0$ such that 
  $\Lambda|_K = \mu|_K$. In fact there are infinitely many such $\Lambda$. Hence, 
  \be
   \left[ Q(\mu), \mathcal{O}_K \right] = \left[ Q(\Lambda|_K), \mathcal{O}_K \right] =  \left[ G(\Lambda|_K), \mathcal{O}_K \right] = \left[ G(\Lambda), \mathcal{O}_K \right]
   \label{trick}
   \ee
where the last step folows from the fact that the commutator depends only on $\Lambda|_K$. Hence, 
 \be
 \left[ Q(\mu), \mathcal{O}_K \right] = 0.
   \ee
   As the local observables commute with $Q(\mu)$, they are all constructed from electric and magnetic fields. Thus $W(x)$ is {\em not} a local observable.

   The important result explained in the last few paragraphs is that observables, being local, commute with both small and large gauge transformations.

    \section{The Self-Dual Action for Gravity}
    The material in this section is  rather well-known \cite{Ashtekar}. It  is added  here for completeness. 
    
    The self-dual gravity action in four dimensions is  
    \be
    S = \int d^4x~ ( det ~e) e_C^a~ e_D^b ~\mathcal{F}_{ab}^{CD} ( A) 
    \label{sdual}
 \ee
 where the upper case $C,D\in {1,2,3,4}$ are Lorentz group indices and $a,b$ are spacetime indices. 
 
 The connection $A$ and its curvature $\mathcal{F}$ are Lorentz Lie algebra-valued, in the self-dual $(1,0)$ representation of the Lie algebra.
 They are complex.  (The antisymmetric product of two four-vector representations contains  $(1,0)$ of course). 
    
For the Palatini action in its real form, $A$ and $\mathcal{F}$ are real in (\ref{sdual}) . The  connection and the curvature are valued  in the four-vector 
representation of the $SL(2,\mathbb{C}) $ Lie algebra.

As mentioned earlier, (\ref{sdual}) has redundant degress of freedom associated with $e^a_A$. This is accompanied by $SL(2, \mathbb{C})$ - gauge invariance. 

After Legendre transformation and associated manipulations, one finds the canonically conjugate variables $A^C_i, \tilde{E}^j_D$ \cite{AshJo}:

\bea
 \left[ A^C_i(x), \tilde{E}^j_D\right] =  \delta^C_D \delta^j_i \delta^3(\mathbf{x} - \mathbf{y}) \\
 \label{sd2}
 \tilde{E}^A _i= \frac{1}{2} \epsilon^{ABC} \epsilon_{ijk} e^j_B e^k_C  \Leftrightarrow  e^i_A = \frac{1}{2} \frac{1}{ | \det \tilde{E}|^\frac{1}{2} }  \epsilon_{ABC} \epsilon^{ijk} \tilde{E}_j^B \tilde{E}_k^C
 \label{sd3}
 \eea
at a fixed time slice. All other commutators involving $A$ and $\tilde{E}$ vanish on this slice. 

Let $T(C)$ be the $SO(3)$ Lie algebra generators in the spin 1 representation: 
\be
T(C)_{AB} = - \epsilon_{CAB}  
\ee
 and set 
\be
A_i = A_i^C T(C), \qquad \tilde{E}^j = \tilde{E}^j_D T(D).
\ee
Then $\mathcal{F}_{ij}(A) = {F}_{ij}^C (A) T(C)$ is also  Lie algebra valued, $F_{ij}^C(A) T(C)$ being the curvature of the connection $A_i$. 

For the self-dual Lagrangian,there are also the following first class constraints in the absence of matter fields: 

\begin{enumerate}[(i)]
 \item The Gauss law: 
 \be
 \mathcal{D}_{i} \tilde{E}^{i} | \cdot \rangle = 0
 \ee
 on allowed vector state vectors $|\cdot \rangle$. Here we have used a standard notation:
 \be 
 \mathcal{D}_i \tilde{E}^i = \partial_i \tilde{E}^i +  A_i^\alpha  \tilde{E}^{i \beta} [ T(\alpha) , T(\beta) ]. 
 \ee

\item The vector constaint: 
\be
\Tr \tilde{E}^i F_{ij} | \cdot \rangle = 0 .
\ee

\item The scalar constraint: 

\be
\Tr \tilde{E}^i \tilde{E}^j F_{ij} | \cdot \rangle = 0.
\ee

\end{enumerate}

\section{ The Self-Dual Action: Edge States}

We focus on the Gauss law. We must smear it with test functions as before. For this purpose, we let $\Lambda$ and $\mu$ henceforth
be Lie algebra valued, $\Lambda \equiv \Lambda^C T(C), \mu = \mu^C T(C)$ where the functions $\Lambda^C \in \mathcal{C}^\infty_0$ while 
$\mu^C \in C^\infty$, i.e. the $\mu^C$ can approach angle dependent limits at infinity. Then with 
\be
G(\Lambda) = - \int d^3x ~ \Tr ( \mathcal{D}_i \Lambda) \tilde{E}^i , 
\ee
one has 
\be
G(\Lambda) | \cdot \rangle = 0 
\ee
which are the Gauss law constraints.

We can also define $Q(\mu)$: 
\be
Q(\mu) = - \int d^3x ~ \Tr ( \mathcal{D}_i \mu) \tilde{E}^i . 
\label{new}
\ee
As mentioned, we now allow the possibility that $\mu^c(x)$ approach angle-dependent limits $\mu^c_\infty(\hat{n})$, as infinity
is approached, as in the abelian case. 

In what follows, $\mu(\hat{n}) $  denotes $\mu^c(\hat{n}) T(c)$. 

The group that $G(\Lambda)$'s generate is identified with $\mathcal{G}^\infty_0$, the group of maps from $\mathbb{R}^3$ to complexified $SO(3)$ 
which become identity maps at $\infty$ and are also connected to identity. The group that $Q(\mu)$'s  generate is instead $\mathcal{G}_0$, which are maps 
from $\mathbb{R}^3$ to complexified $SO(3)$, whose 
elements may not approach identity at $\infty$, but are connected to identity. 

If $g$ is an element of $\mathcal{G}_0$ or $\mathcal{G}^\infty_0$, we denote its representative element in 
quantum theory by $U(g)$. 

On quantum states, the effective group is $\mathcal{G}_0 / \mathcal{G}^\infty_0$. Its representations give the edge states of the self-dual 
action. These are edge states since their action on quantum fields only depends on the asymptotic data of $S^2_\infty$. 
In particular, in analogy to the $U(1)$ case, if $\mu_\infty$ are restricted to constant functions, $Q(\mu)$ generate the 
complex $SO(3)$ Lie algebra while  if  $\mu_\infty$ is allowed to have $\hat{n}$-dependence, we get the Sky group or a subgroup thereof.
This Sky group is a natural generalisation of the one formulated for gauge theories \cite{Bal_2} to the self-dual Palatini action. 

To understand this material better, we generalise (\ref{sky}) and (\ref{trick}).  Instead of $\Psi_0$ we use $\tilde{E}^i$. Then 
(\ref{phase1}) becomes 
\be
  U(e^{i\Lambda})  \tilde{E}^i (x)  U(e^{-i\Lambda}) = e^{ i  \Lambda(x)} \tilde{E}^i (x)
\label{Pal_Sky}
\ee
where now $\Lambda(x)$ is a matrix like $T(\alpha)$  while $E^{~i}$  is a column vector (  $E^{~i}_C $ ) in 
the adjoint representation  and not its contraction with $T(C)$ as in section 3. 
Thus we are changing notation for convenience.

The Wilson line $W(x)$ looks the same as (\ref{Wil_line}), except that it now includes a path-ordering $\mathcal{P}$: 
\be
W(x) = \mathcal{P} \exp{i \int_x^\infty dx^\lambda A_\lambda(x)} 
\label{Wil_2}
\ee
\be
  U(e^{ i\mu})  W(x)  U(e^{-i\mu}) = e^{ i  \mu_\infty ( \hat{n} ) } W(x) e^{-i \mu(x)} 
  \label{Wil_3}
 \ee

In (\ref{sky}), we change $\Psi(x)$ to 
\be
W(x) \tilde{E}^i(x) 
\label{change}
\ee
so that 
\be
  U(e^{i \mu}) W(x) \tilde{E}^i (x) U( e^{-i\mu}) = e^{ i  \mu_\infty ( \hat{n} ) } W(x) \tilde{E}^i(x) .
\ee
Note that $\mu_\infty (\hat{n})$ is a matrix. We can create vector states characterised by $\mu_\infty(\hat{n})$ 
in its transformation 
law by starting with a vector $| \cdot \rangle$ invariant under $U(e^{i\mu})$: 
\be 
U(e^{i\mu}) | \cdot \rangle = | \cdot \rangle.
\ee
Then  applying (\ref{change}) on the vacuum, we get such a state since one gets  
\be
U(e^{i\mu}) W(x) \tilde{E}^i(x) | 0 \rangle = U(e^{i\mu_\infty( \hat{n})}) W(x) \tilde{E}^i(x) | \cdot \rangle .
\label{mixing} 
\ee
From this equation, we read off that if $\mu(\hat{n}) $ runs over all constant values, we get the complexified 
$SO(3)$ whereas if $\mu_\infty$ runs over all functions on $S^2_\infty$, we get the complexified 
Sky group. 

\subsection*{Remarks on Unitarity}

The transformation property (\ref{Wil_3}) of $W(x)$ depends only on $A$ transforming as a connection and not on its reality. 
Loop quantum gravity program \cite{Ashtekar} extensively employs this fact. But $W(x)$ is not even formally unitary since $A_i$ are complex. 
The implications of this observation are not clear.

\section{ On Observables and Superselection Rules} 

Locality is a concept which is difficult to formulate in quantum gravity, as the concept has to be diffeomorphism invariant. 
Because of this requirement, no  useful diffeomorphism invariant {\em local} observables have been found on $\mathbb{R}^3$.

Besides this issue, we have the requirement that whatever be the definition of observables, they must commute with $G(\Lambda)$. 
Now $\tilde{E}^a_i$ and $e_a^i$ undergo $SL(2, \mathbb{C})$ - gauge transformations under the action of $U(e^{i\Lambda})$ . 
For example,  
\be
U(e^{i\Lambda}) \tilde{E}_i^a(x) U(e^{-i\Lambda})  = \left(e^{i\Lambda(x)}\right)^a_{~~b} \tilde{E}^b_i (x)
\label{trans}
\ee
where $e^{i\Lambda}$  become identity as $x \rightarrow \infty$. Invariance of observables under (\ref{trans}) means 
that no observable can depend on $\tilde{E}^a_i$  or $e^i_a$ for finite $x$  except in $SL(2, \mathbb{C})$ invariant 
combinations. 
But they can depend on the frames ``at infinity'', the asymptotic forms of $e^i_a(x)$, as $x \rightarrow \infty$. 

 In QED, because of locality, local observables commuted with $Q(\mu)$ due to locality\cite{Bal_2} since they did so with $G(\Lambda)$ .
 But that argument is not available for the case of gravity.
 
 But suppose that we want the equivalence of the self-dual gravity at the level of observables with the 
 Einstein-Hilbert action. Then we must 
 assume that all its observables $\mathcal{O}$ commute with $Q(\mu)$. To proceed, let us agree on this suggestion. 
 Then if $\mathcal{O}$ is any observable, (\ref{mixing}) 
 shows that 
 \be
U(e^{i\mu}) \mathcal{O} W(x) \tilde{E}^i(x) | 0 \rangle =  U(e^{i\mu_\infty( \hat{n})}) \mathcal{O}W(x) \tilde{E}^i(x) | \cdot \rangle .
\label{tw}
\ee
( Here $\mathcal{O}$ is not a matrix  while $\tilde{E}^{~i}$  is a column vector  
(  $ \tilde{E}^{~i}_b $ ) in the adjoint representation.  ). 
It follows that the collection of numbers $\{ \mu_\infty (\hat{n})\}$ is characteristic of the representation of 
the algebra of observables $\mathcal{O}$. It is thus superselected, as $\mathcal{O}$ does not affect this set.  
[ We note that if $U(e^{i\mu}) \rightarrow U(e^{i \mu_\infty})$
in (\ref{tw}) , and similarly   $U(e^{i\mu' })\rightarrow U(e^{i \mu'_\infty}) $, then 
$U(e^{i\mu})U( e^{i \mu'})  \rightarrow U(e^{i \mu'_\infty})U(e^{i \mu_\infty})$, which 
shows an inversion of the usual ordering. ( Just as for $U(e^{i\mu}), U(e^{i\mu_\infty})$ and similar expressions represent the quantum operators for their arguments.) ] 

We can change $\{ \mu_\infty ( \hat{n} ) \}$ to  $\{ \mu_\infty ( \hat{n}' ) \}$ by changing the direction of 
the Wilson line and choosing the latter to be 
\be
\{ x + \hat{n}'  l , \quad 0 \leq l < \infty \} .
\label{direc2}
\ee
Since by (\ref{trans}), no observable can change  $\{ \mu_\infty ( \hat{n} ) \}$ to  $\{ \mu_\infty ( \hat{n'} ) \}$ , they are different representations of 
$\mathcal{A}$. 

Thus we have an infinity of representations characterised by $\hat{n}$, that is, points on the celestial sphere $S^2_\infty$.  They are absent for the 
observables coming from the Einstein-Hilbert action which does not use the frames. 

\section{Mixed States in Self-Dual Gravity?} 

Let $A_i, \tilde{E}^i$ represent a particular representation of these fields as conjugate pairs. Then a large gauge transformation $e^{iQ(\mu)} $ transforms them to another 
equivalent representation ( which however may not be unitarily equivalent!) , namely  $A_i, \tilde{E}^i$ to $  e^{iQ(\mu)}A_i  e^{-iQ(\mu)}$ and  $ e^{iQ(\mu)}\tilde{E}^i e^{-iQ(\mu)}$. 
Hence also $W$ is changed to $ e^{iQ(\mu)}W  e^{-iQ(\mu)}$ . Thus the new state vector, presumably equivalent in some sense to $W(x)\tilde{E}^i(x) |  \cdot \rangle$ is, 
\be
 U(e^ {i\mu}) W(x) \tilde{E}^i  | \cdot \rangle  =  e^{i \mu_\infty(\hat{n}) } W(x) \tilde{E}^i (x) |\cdot \rangle
 \label{6.1}
 \ee
 since  $| \cdot \rangle $ is invariant by $U(e^{i\mu})$ by the assumption in Sec. 5.(Cf.(4.9).)

 Observables commute with $U(e^{i\mu})$. But still matrix elements of observables between vectors  like
 (\ref{6.1} ) can depend 
 on $U(e^{i\mu}) $ if it is not unitary. We shall henceforth avoid this problem by considering only $SO(3)$ - 
 ( and not $SL(2,\mathbb{C})$-)valued 
 gauge transformations. In that case, $U(e^{i  \mu })$ being unitary , 
 matrix elements of $\mathcal{O}$ 
 between vectors (\ref{6.1}),
 are not expected to depend on $e^{i \mu_\infty(\hat{n})}  $.  
 
 By considering $U(e^{i\nu })U( e^{i\mu}) $ etc. in (\ref{6.1}) where also  $e^{i\nu_\infty( \hat{n}) } \in SO(3)$, 
 we get a family of vectors in (\ref{6.1}) giving equal matrix elements for observables  $\mathcal{O}$, 
 parametrised by maps of $S^2_\infty$ to $SO(3)$. 
 
 Consider the density matrices 
 \be
  \rho (\mu ) =  U(e^{i\mu}) W_a^{~b}(x) \tilde{E}^i_{~b}(x)| \cdot \rangle \langle \cdot | (\tilde{E}_{~c}^i)^\dagger {W^{-1}}^c_{~a}(x)  
  U(e^{-i\mu}). 
  \label{6.2}
 \ee
 for fixed $a$. 
By the preceding considerations, for this $\rho$, the expectation value of  $\mathcal{O} \in \mathcal{A}$ is independent of $\mu$. That is
also the case for convex 
linear combinations 
\be
\sum \lambda_i  \rho(\hat{\mu}^{(i)} ) , \qquad \lambda_i \geq 0 , \sum_i \lambda_i = 1
\label{6.3}
\ee
of such $\rho(\hat{\mu}^{(i)})$.  That means that $\rho(\hat{\mu})$ is a mixed state, , having many decompositions (\ref{6.3}).

We can generalise (\ref{6.3}) by replacing $U(e^{\pm i\mu})$ in (\ref{6.2}) by $U(e^{\pm i \nu} ) U( e^{\pm i\mu}) $ . In this way, we get the 
decomposition of $\rho(\mu ) $ as the convex 
sums of  many states. 

The vector $| \cdot \rangle $ can be any vector invariant by the action of $U(e^{i \Lambda})$, that is , a small gauge transformation. It can 
in particular be an $SO(3)$ singlet.
Still (\ref{6.2}) depends on $\mu$ since the remaining factors do not commute with $U(e^{ i \mu})$.Thus (6.3) depends on $\lambda_i$. 

We thus find that the above $SO(3)$ non-singlet states are mixed.

There is a similar result for {\em all} $SO(3)$ non-singlet states. For this purpose, consider a generic $SO(3)$  
non-singlet state $| \psi, \cdot \rangle$ normalised to unity and fulfilling $ G( \Lambda) | \psi \rangle=0$. 
The transformations of $SO(3)$ are implemented by $U(e^{i\mu })$ where $\mu_\infty$ is a constant on $S^2_\infty$. 
Since $SO(3)$ is compact, we can reduce its action to a direct sum of irreducible representations. Hence we can write 
\be
| \psi  \rangle = \sum C_{I I_3 \lambda } | I, I_3, \lambda \rangle , \qquad I \in \{0,1,2, \cdots\} , \qquad I_3 \in [ -I, +I]. 
\label{6.4}
\ee
where $\lambda$ accounts for labels not covered by $I,I_3$ . Also $| I, I_3, \lambda \rangle$ are orthonormal .

The observables do not mix $I$ and $I_3$ as they commute with $Q(\mu)$. 
Assume for a moment  that they do not mix $\lambda$ as well. Then the density matrix $|\psi \rangle \langle \psi|$
gives the same expectation values as 
\be
\sum_{I, I_3, \lambda}  C_{II_3 \lambda} |I, I _3, \lambda \rangle \langle I, I_3, \lambda| \bar{C}_{II_3 \lambda}. 
\label{6.4a}
\ee

Each term in (\ref{6.4a}) with $I \neq 0$ leads to a mixed state. That is, the density matrix 
\be
\rho_{I I_3 \lambda} = | I, I_3, \lambda \rangle \langle  I, I_3, \lambda | 
\label{6.5}
\ee
despite appearance, is mixed  \cite{C2H4}!  Thus for example 
$ \Tr \rho_{I I_3 \lambda} \mathcal{O} = \Tr \rho_{I I'_3 \lambda} \mathcal{O}$, 
if the two $\rho$'s are related by an $SO(3)$-transformation. That is because the latter are generated by $Q(\mu)$ with constant $\mu_\infty$ 
and $Q(\mu)$ commutes with observables by assumption in Sec.5. More generally,  we see that 
as states on observables,  $\rho_{I I_3 \lambda}$ and $ \xi  \rho_{I I_3 \lambda}  + ( 1 - \xi)  \rho_{I I'_3 \lambda}$ 
( $0 \leq \xi \leq 1)$ all define the same state (( give the same expectation values). 
Hence $(\ref{6.4a})$ is mixed. 

In general, $\mathcal{O}$ will mix different $\lambda$'s ,  but the argument is easily adapted to  the case of 
several $\lambda$'s in the density matrix (\ref{6.4a}) , its 
term for fixed $I, I_3$ then being 
\be
\sum_{\lambda, \lambda'} | I, I_3, \lambda \rangle  C_{I, I_3, \lambda} \bar{C}_{I, I_3, \lambda'}  \langle  I, I_3, \lambda' |
\label{6.6}
\ee
where $C_{I,I_3, \lambda}$ are complex numbers. One sees this from (\ref{6.4}).

The mixed states implied by  the existence of the edge states in the self-dual model
are quite interesting. They may give clues for the black hole entropy problem.

\section*{Acknowledgements}
A.P.B. thanks Rakesh Tibrewala for kindly reminding him about  the equation (\ref{sd2}). We are deeply grateful to Giorgio Immirzi 
for very helpful suggestions and correcting errors. After submitting this article to the arXiv, our attention was drawn to the 
reference \cite{marc_g} where edge states in the context of three dimensional gravity is discussed.


\begin{thebibliography}{999}
 \bibitem{Book_1} J. F. Cari\^{n}ena, A. Ibort , G. Marmo  and  G. Morandi- Geometry from Dynamics, Classical and Quantum  ( Springer, 2015) and references therein.  
 \bibitem{Bal_2} A. P. Balachandran, T.R. Govindarajan, B. Vijayalakshmi,  Phys.Rev. {\bf D18} (1978) 1950 and references therein. 
 \bibitem{Bal_Chandar}A. P.  Balachandran, L. Chandar and A. Momen, Nucl.Phys. {\bf B461 }(1996) 581; A. Momen, Phys.Lett. {\bf B394} (1997) 269 and references therein. 
 \bibitem{C2H4} A.P. Balachandran, A.R. de Queiroz, S. Vaidya Eur.Phys.J.Plus {\bf 128} (2013) 112;  Phys.Rev. {\bf D88 }(2013)  025001. See also \cite{Bal_QCD}; A. P. Balachandran and V.P. Nair, 
 {\it in preparation}. 
  \bibitem{Strominger} A. Strominger,  Lectures on the Infrared Structure of Gravity and Gauge Theory  (arXiv:1703.05448 ) and references therein.
 \bibitem{Dirac} P. A. M. Dirac, Can. J. Phys. {\bf 33}(1955) 650.
 \bibitem{Mandelstam} S. Mandelstam, Ann. Phys. NY {\bf 19} (1962) 1. 
 \bibitem{Mix}A.P. Balachandran and  S. Vaidya, Eur.Phys.J.Plus {\bf 128} (2013) 118;   A.P. Balachandran, S. K\"{u}rk\c{c}\"{u}oglu, A.R. de Queiroz, S. Vaidya, Eur.Phys.J. {\bf C75 }(2015), 89.
  \bibitem{Bal_QCD}A. P. Balachandran,  Mod.Phys.Lett. {\bf A31} (2016) 1650060.  
 \bibitem{Ashtekar} A. Ashtekar and R. Tate. Lectures on Non-Perturbative Canonical Gravity (  World Scientific, 1991) and references therein.  
 \bibitem{AshJo} A. Ashtekar,  A.P. Balachandran, S. Jo,  Int.J.Mod.Phys. {\bf  A4} (1989) 1493 .
 \bibitem{marc_g} M. Geiller, JHEP 02 (2018) 29.
 \end{thebibliography}
\end{document}